\documentclass[preprint,12pt]{aastex}
\usepackage{amsmath}

\title{Cluster Morphologies as a Test of Different Cosmological Models}
\author{Tamon Suwa and Asao Habe}
\affil{Division of Physics, Graduate School of Science, Hokkaido University,
	Sapporo 060-0810, Japan}
\email{tamon@astro1.sci.hokudai.ac.jp}
\email{habe@astro1.sci.hokudai.ac.jp}
\author{Kohji Yoshikawa}
\affil{Department of Astronomy, Kyoto University, Kyoto 606-8502, Japan}
\email{kohji@kusastro.kyoto-u.ac.jp}
\and
\author{Takashi Okamoto}
\affil{Yukawa Institute for Theoretical Physics, Kyoto University, Kyoto 606-8502, Japan\\
AND\\
Department of Physics, University of Durham, South Road, Durham, DH1 3LE, England}
\email{Takashi.Okamoto@durham.ac.uk}

\shorttitle{Clusters morphologies as a Test of Cosmology}
\shortauthors{Suwa et al.}

\newcommand{\Mint}{M_{\mathrm{int}}}
\newcommand{\bx}{\boldsymbol{x}}

\begin{document}

\begin{abstract}
We investigate how cluster morphology is affected by the cosmological constant 
in low-density universes. 
Using high-resolution cosmological N-body/SPH simulations 
of flat ($\Omega_0 = 0.3, \ \lambda_0 = 0.7, \Lambda$CDM) 
and open ($\Omega_0 = 0.3, \lambda_0 = 0,$ OCDM) cold dark matter 
universes, we calculate statistical indicators to quantify 
the irregularity of the cluster morphologies.  
We study axial ratios, center shifts, cluster 
clumpiness, and multipole moment power ratios as indicators 
for the simulated clusters at $z=0$ and $0.5$.
Some of these indicators are calculated for both the X-ray surface brightness 
and projected mass distributions. 
In $\Lambda$CDM all these indicators tend to be larger than those
in OCDM at $z=0$.
This result is consistent with the analytical prediction of Richstone, 
Loeb, \& Turner, that is, clusters in $\Lambda$CDM are formed later than in 
OCDM, and have more substructure at $z=0$.
We make a Kolmogorov-Smirnov test on each indicator for these two models.
We then find that the results for the multipole moment power ratios and 
the center shifts for the X-ray surface brightness 
are under the significance level (5\%).
We results also show that these two cosmological models can be distinguished 
more clearly at $z=0$ than $z = 0.5$ by these indicators.
\end{abstract}

\section{INTRODUCTION}
Measurement of the cosmological parameters, such as the Hubble constant $H_0$,
the density parameter $\Omega_0$, and the cosmological constant
$\Lambda$, is one of the most important studies in observational 
cosmology.
From observational evidence, the density parameter $\Omega_0$ is estimated
to be around 0.3 \citep[e.g.][]{Bahcall99}.
The inflationary cosmology requires $\Omega_0 + \lambda_0 = 1$, where 
$\lambda_0=\Lambda/(3H_0^2)$, and  
is supported by recent observations of the cosmic microwave background 
\citep[e.g.][]{Boomerang}.
Distant Type Ia supernovae (SNe Ia) are assumed to be standard candles
to explore the effect of a cosmological constant by searching for evidence of 
accelerated expansion \citep{SCP,HZS}. Several systematic
uncertainties in the observational data of
SNe Ia have been pointed out \citep{TK99,Recalibration}.
Therefore, independent studies in the non-linear regime are important to
confirm the values of the cosmological parameters. 

Richstone, Loeb, \& Turner (1992) proposed that the fraction of clusters
which have significant substructure 
can be a probe of the density parameter, $\Omega_0$, since the fraction of 
recently formed clusters strongly depends upon $\Omega_0$ and 
such young clusters are likely to have substructure.
They have also revealed that the formation epoch of clusters in 
a universe with a cosmological constant is later than that in a universe 
without a cosmological constant, although the effect of the cosmological 
constant is much smaller than that of the density parameter.
In fact, Wilson, Cole, \& Frenk (1996) showed that the statistics of 
the quadruples of the column mass density of clusters strongly depends 
on the value of $\Omega_0$, 
while it is quite insensitive to the value of $\lambda_0$. 

Some authors have studied if cosmological models with and without 
a cosmological constant can be discriminated by the statistics of 
the irregularity of cluster morphologies 
(Jing et al. 1995; Crone, Evrard, \& Richstone 1996; Buote \& Xu 1997).
They performed pure N-body simulations in low-density flat universes 
($\Omega_0=0.2$--$0.35,\, \lambda_0 = 1-\Omega_0$) and open universes 
($\Omega_0=0.2$--$0.35,\, \lambda_0=0$).
\citet{Jing et al.95} and \citet{CER96} quantified morphologies of X-ray 
surface brightness of clusters from the $N$-body results assuming 
hydrostatic equilibrium of the hot gas in the gravitational potentials of 
the clusters.
Using a Kolmogorov-Smirnov test (hereafter KS-test), they showed that 
a low-density flat universe and an open universe are distinguishable 
from each other by the center shifts.
On the other hand, \citet{BX97} obtained multipole moment power ratios 
for clusters in their N-body simulations and compared those in 
a low-density flat universe and an open universe.
They concluded that these two cosmological models are not
distinguishable by the multipole power ratios. 
Since these authors estimated the hot gas distribution from N-body 
simulations, their results are in question, and then simulations
including hydrodynamics are desired. 

In previous hydrodynamic simulations \citep{Evrard93,M95} the numbers of 
clusters were not enough to perform a statistical comparison. 
Although Valdarnini, Ghizzardi, \& Bonometto (1999) analyzed a
sufficient number of clusters for various cosmological models, 
the effect of the tidal field 
from larger scales ($\gtrsim 30$--$50 h^{-1}$Mpc) was neglected in their 
simulations because they performed 
tree-SPH simulations of individual clusters selected from the large $N$-body 
simulations. 
By comparing the simulated clusters and the ROSAT X-ray cluster data, 
they pointed out that the statistical test for the observed data favors 
the high-density ($\Omega_0=1$, SCDM) model rather than the \textit{concordant} 
low-density flat ($\Omega_0=0.3,\lambda_0=0.7$) model.
It is worthwhile to confirm their results by simulations that  
take account of the tidal field from outside the clusters. 
We then simulate comoving 150 $h^{-1}\mathrm{Mpc}$ boxes with 
particle-particle-particle-mesh ($\mathrm{P^3M}$)-Smoothed Particle 
Hydrodynamics (SPH) code. 

In this paper, we study morphologies of simulated clusters in flat 
($\Omega_0 = 0.3,\, \lambda_0 = 0.7,\, \Lambda$CDM) and open 
($\Omega_0 = 0.3,\, \lambda_0 = 0$, OCDM) cold dark matter universes 
by using high resolution simulations.
We perform large simulations to obtain a sufficient number of clusters 
for statistical comparisons.
A reason why we study the difference between low-density universes 
despite the fact that recent CMB data  \citep[e.g.][]{Boomerang} supports 
a flat universe is as follows.
As we mentioned above, the irregularity of cluster morphologies is 
quite sensitive to the value of $\Omega_0$ while it is rather insensitive 
to the value of $\lambda_0$. 
Then the statistical indicators that can discriminate between the 
low-density models with and without the cosmological constant $\lambda_0$ 
enable us to detect the small difference in the formation histories of 
clusters in different but similar cosmologies such as the low-density 
universe with a time dependent vacuum energy. 
Furthermore, because the CMB data provide us only the information about 
the linear density fluctuations, the consistency test in the non-linear 
regime is very important. 
For instance, if the finding of \citet{Valdarnini99} is true, 
we should consider what causes the discrepancy. 
Unfortunately, the resolution in our simulations is not sufficient to 
make a direct comparison with high-resolution X-ray observations by Chandra or 
XMM and our simulations do not include some important physical processes 
like the cooling and heating, which may affect the morphology of the 
X-ray surface brightness. 
Therefore here we restrict ourselves to finding effective 
statistical indicators to discriminate between two low-density universes 
and we do not compare our simulation results with the observations.

We identify clusters in our simulation and calculate various statistical 
indicators for them.
The KS-tests are performed to measure how effectively these indicators 
distinguish between these two cosmological models. 
These indicators are calculated for the projected mass density as well as 
the X-ray surface brightness of clusters of galaxies. 
The X-ray surface brightness reflects the distribution of the hot gas in 
the cluster. 
On the other hand, the projected mass density is dominated by the distribution 
of dark matter. 
Although dark matter cannot be observed directly, a method by which 
the mass distribution is reconstructed from small distortion images of 
background galaxies caused by gravitational lensing (these small distortions 
are called the `weak shear field') has been developed recently 
\citep[e.g.][]{AR99,LK97,Clowe00}.
The precision of this method is not yet sufficient to probe cluster 
substructure. 
It is, however, likely that this mass reconstruction method will progress and
become a useful tool to measure substructure in galaxy clusters.
\citet{SB97} have already investigated the possibility of deriving multipole 
moments of the projected mass distribution of clusters with weak-lensing.  

The organization of this paper is as follows.
In \S\ref{sec_method}, we describe the numerical simulations used in this 
paper, the method of cluster identification, and the definitions of indicators 
which we use.
We give mean values and standard deviations of the indicators for simulated clusters and show 
results of the KS-tests on these indicators in \S\ref{sec_results}.
In \S\ref{sec_discussion}, we discuss our results and present our conclusion.

\section{METHOD} \label{sec_method}
\subsection{Numerical Simulations} \label{sec_data}
Our simulations are based on a $\mathrm{P^3M}$-SPH algorithm.
Detailed description of our simulation method is given in Yoshikawa, Jing, \& 
Suto (2000).

We use two cosmological models, OCDM and $\Lambda$CDM.
Cosmological parameters in the two models, except for the cosmological 
constant, are as follows: the Hubble 
constant in units of $100 \mathrm{km\,s^{-1}\,Mpc^{-1}}$, $h=0.7$, 
the density parameter, $\Omega_0 = 0.3$, the baryon density parameter, 
$\Omega_b = 0.015h^{-2}$, the rms density fluctuation amplitude on a scale 
$8h^{-1}\mathrm{Mpc}$, $\sigma_8=1.0$, and the power-law index of the 
primordial density fluctuations, $n=1.0$.
The normalized cosmological constant, $\lambda_0$ is 0 and 0.7 for OCDM 
and $\Lambda$CDM, respectively. 
The simulation of $\Lambda$CDM is the same as L150A in \citet{YJS2000} 
and that of OCDM is carried out for this paper.

Each simulation employs $N_{\mathrm{DM}}=128^3$ dark matter particles 
and the same number of SPH particles.
The mass of a dark matter particle and an SPH particle are $1.7\times 
10^{11}M_\odot$ and $2.0\times 10^{10}M_\odot$, respectively.
The size of comoving simulation box, $L_{\mathrm{box}}$, is 
$150h^{-1}\mathrm{Mpc}$, and the box has the periodic boundary
conditions.
We use the spline (S2) gravitational softening 
\citep{softening}, and the softening length, $\epsilon_{\mathrm{grav}}$, 
is set to be $L_{\mathrm{box}}/(10N_{\mathrm{DM}}^{1/3})$ 
($\sim 120h^{-1}\mathrm{kpc}$).
The smoothing length of each SPH particle is determined by
\begin{equation}
h^{(n)}_{i} = \frac{1}{2}\left\{ 1+ \left(\frac{N_s}{N_i^{(n-1)}}\right)^{1/3}\right\} h^{(n-1)}_{i},\label{eq_smoothing}
\end{equation}
where $h^{(n)}_i$ is the smoothing length of the $i$-th particle at the $n$-th 
time step, $N_i^{(n)}$ is the number of neighbor particles of the $i$-th 
particle at the $n$-th time step (the number of particles inside a 
sphere of radius $h^{(n)}_i$), and $N_s$ is mean number of neighbor 
particles \citep{HK89}.
We set $N_s = 32$ and the minimum SPH smoothing length as $h_{\mathrm{min}} = 
\epsilon_{\mathrm{grav}}/4$ ($\sim 30h^{-1}\mathrm{kpc}$).
We use the COSMICS package \citep{Bert95} to generate initial conditions
at $z=25$.
Our simulations are carried out on a VPP5000 and each run took about 
50 hours using 8PE with parallel $\mathrm{P^3M}$-SPH.
The memory resource needed to run each simulation is 500MB.

\begin{figure}
	\plotone{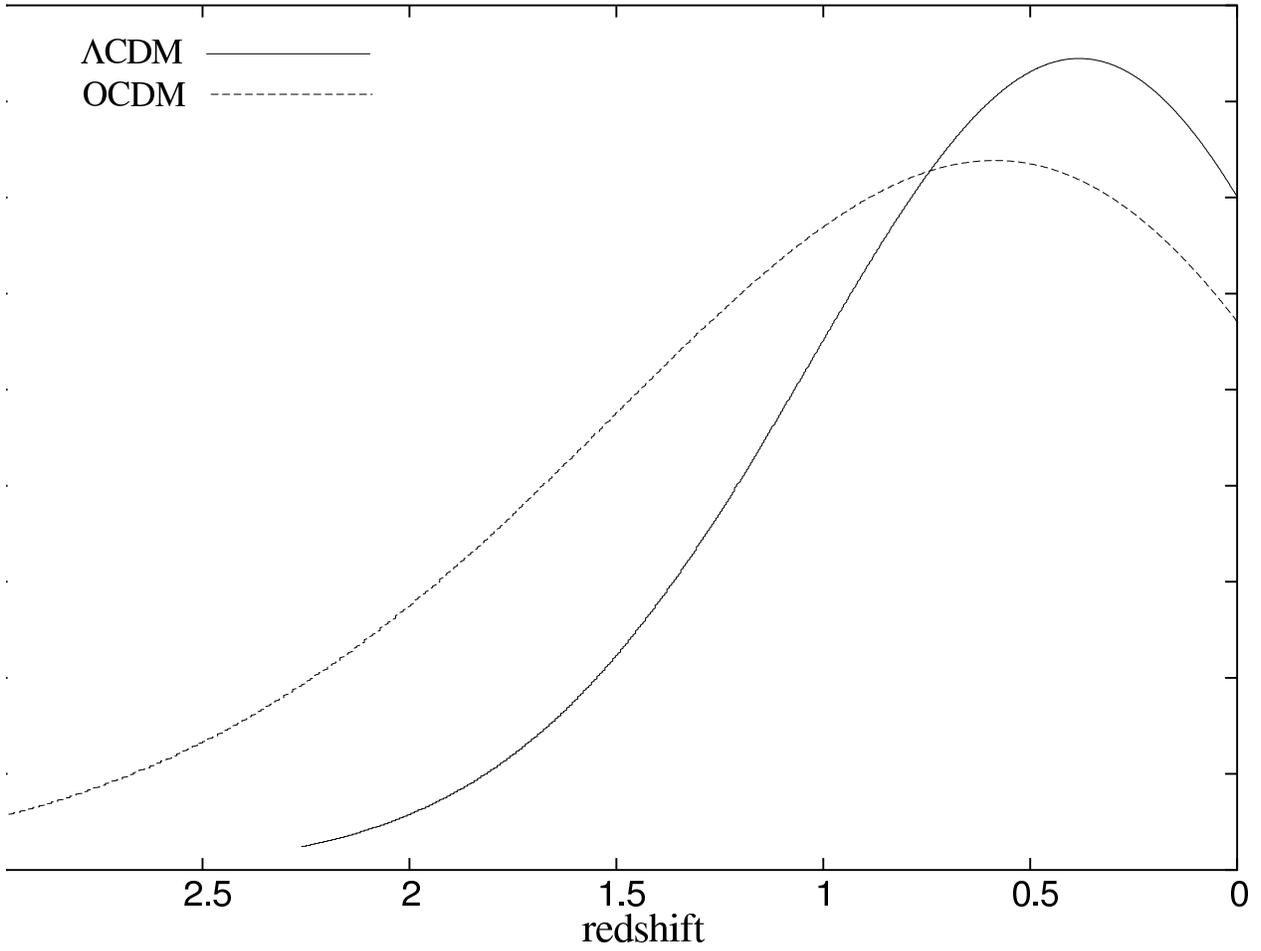}
	\caption{The formation rate of galaxy clusters.
		The solid curve indicates the formation rate in $\Lambda$CDM
		and the dashed curve indicates that in OCDM.}
	\label{fig_rate}
\end{figure}

We calculate statistical indicators for the clusters in our simulations at 
$z=0.5$ and $0$.
The reason for this is that the formation rate of galaxy clusters in
$\Lambda$CDM is expected to exceed that in OCDM for $z<0.8$ as shown in
Fig.\ref{fig_rate} which shows the formation rate of galaxy clusters as
a function of redshift according to the analytical formula given by 
\citet{RLT92}.
In Fig.\ref{fig_rate} the number of formed clusters is normalized 
by the present value and the cosmic time is also normalized by the present 
age of the universe.
The solid and dashed curves indicate the formation rates in $\Lambda$CDM and
in OCDM, respectively.
Since just before $z=0.5$ the formation rate in $\Lambda$CDM exceeds
that in OCDM, we expect that the difference due to the cosmological
constant can already be detected at $z=0.5$.
Moreover, mass reconstruction based on observations of the 
week shear field has an advantage for high-redshift clusters rather than 
nearby clusters in principle.

\subsection{Cluster Identification} \label{sec_identify}
The way of identifying galaxy clusters is as follows \citep{JF94,Thomas et al.98}.
\begin{enumerate}
	\item We pick up SPH particles whose gas densities are more than 200 
		times the background (gas and dark matter) density.
		Here, we define `gas density of an SPH particle' as gas density at 
		the position of the SPH particle.
		Processes at step 2 and 3 are performed for these selected particles.
	\item We perform a Friends-of-Friends method \citep[hereafter FoF; e.g.][]{Davis85} 
		for the dense SPH particles selected by step 1.
		The linking length, $l$, is defined as $b\bar{n}^{-1/3}$, where $\bar{n}$
		is mean number density of all SPH and dark matter particles in the 
		simulation box, and the constant parameter $b=0.5$.
		It should be noted that the dense SPH particles are confined within 
		small scale areas.
		Since the typical distance between these areas is significantly larger 
		than $l\,(=b\bar{n}^{-1/3})$, the number of groups found by FoF 
		does not depend on $b$ strongly.
		We confirm that both the number of the groups found in this step and 
		the number of clusters found in following steps are not so different 
		for $0.2 \leq b \leq 1.0$.
	\item The densest SPH particle in each group is defined as a `core'
		particle of the group.
	\item We draw a sphere of which the center is on the position of the `core', 
		and seek a radius in which the mean density of total matter is 200 times 
		the background density of total matter.
		This radius is called $r_{200}$.
	\item If the total mass in the sphere with radius $r_{200}$ is more than
		$2\times 10^{14}h^{-1}M_\sun$, a set of particles in the sphere
		is a candidate for being a cluster of galaxies.
	\item If another `core' particle exists in the sphere, the set which
		belongs to the less dense `core' is removed from the cluster candidate
		list.
\end{enumerate}
The candidates which are not removed finally are identified as clusters of
galaxies.
We define the center of a cluster by the position of the `core' particle.
We find that this position is very close to the particle having the minimum 
gravitational potential in the cluster and the distance between the 
`core' particle and the potential minimum is at most several ten kpc 
in our simulation. 
We also find that the position of the `core' is close to the center of mass 
of the cluster except for some clusters which have large substructures.

\begin{figure}
	\epsscale{0.8}
	\plotone{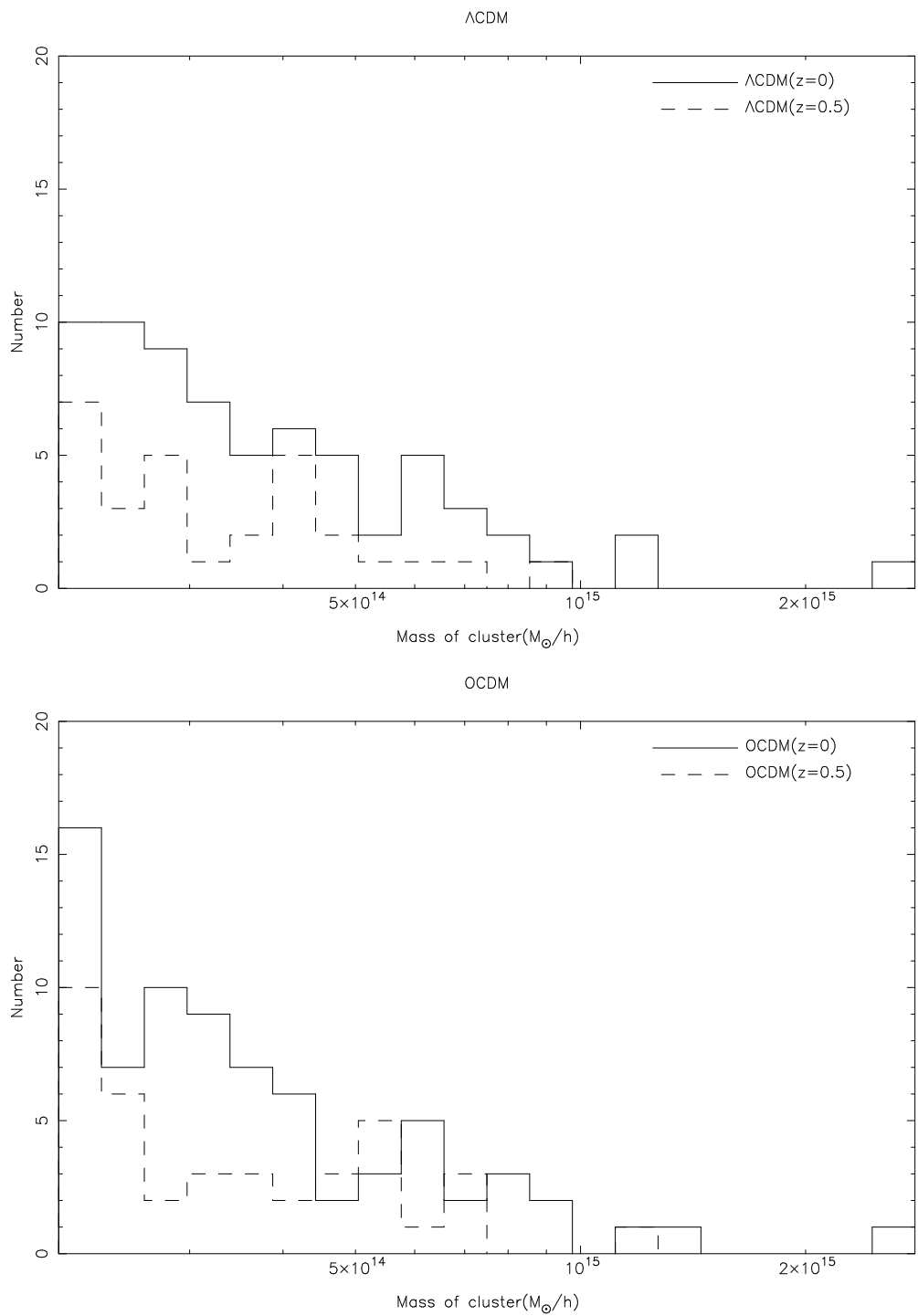}
	\caption{Mass distribution of the clusters.
		The upper panel describes the distribution in $\Lambda$CDM and 
		the lower one describes the distribution in OCDM.
		The solid line indicates the number of clusters at $z=0$ and 
		the dashed line indicates the number at $z=0.5$.
	}
	\label{fig_nM}
\end{figure}

Fig.\ref{fig_nM} shows the mass distribution of the clusters.
The upper panel describes the distribution in $\Lambda$CDM and 
the lower one describes the distribution in OCDM.
The solid line indicates the number of clusters at $z=0$ and 
the dashed line indicates the number at $z=0.5$.
The maximum masses of clusters are $2.7\times10^{15} h^{-1}M_{\sun}$ 
($\Lambda$CDM, $z=0$), $9.3\times 10^{14}h^{-1}M_{\sun}$ 
($\Lambda$CDM, $z=0.5$), $2.7\times 10^{15}h^{-1}M_{\sun}$ 
(OCDM, $z=0$), and $1.2\times 10^{15} h^{-1}M_{\sun}$ (OCDM, $z=0.5$). 
Table \ref{tab_numOfClusters} shows the number of clusters found in each model.
The numbers of clusters increase from $z=0.5$ to $z=0$.
Table \ref{tab_massratioOfClusters} is the ratio of the mass inside clusters
to the total mass in the simulation box.
These ratios in both models are similar to each other at $z=0$, but
the ratio in $\Lambda$CDM is smaller than that in OCDM at $z=0.5$.
This is consistent with the fact that $\Lambda$CDM clusters form later than 
OCDM clusters.
\begin{table}
\begin{center}
	\begin{tabular}{lrr}
	\tableline\tableline
	Model & z=0 & z=0.5 \\\tableline
	$\Lambda$CDM & 66 & 28\\
	OCDM & 74 & 38 \\\tableline
	\end{tabular}
\end{center}
	\caption{The number of the clusters for each model and redshift.}
	\label{tab_numOfClusters}
\end{table}
\begin{table}
\begin{center}
	\begin{tabular}{lrr}
	\tableline\tableline
	Model & z=0 & z=0.5 \\\tableline
	$\Lambda$CDM & 10.2\% & 3.64\% \\
	OCDM & 11.4\% & 5.21\% \\\tableline
	\end{tabular}
\end{center}
	\caption{The ratios of the mass involved in the clusters to total mass in each simulation.}
	\label{tab_massratioOfClusters}
\end{table}

\subsection{Calculation of Column Density and X-ray Surface Brightness} \label{sec_projection}
Two of the indicators we use in this paper, the center shifts and
the power ratios, are calculated for X-ray surface brightness
and the projected mass density for simulated clusters.
Our numerical projection method from 3D to 2D is as follows.

At first, for each cluster, we prepare a cube centered on a core particle 
and the sides of the cube are set to $2r_{200}$.
The total number of grid points in the cube is $128^3$.
Each grid point is assigned indices $(i,j,k)$ and this point is
expressed by $\boldsymbol{r}_{ijk}$.
We calculate gas density and temperature at each grid point
from SPH particles in this cube.
The gas density, $\rho_{\mathrm{gas}}$, and temperature, $T$, at a grid point 
$\boldsymbol{r}_{ijk}$ are given by
\begin{align}
\rho_{\mathrm{gas}}(\boldsymbol{r}_{ijk}) &= \sum_s W(r_{s;ijk}/h_s) \frac{m_s}{h_s^3}, \\
T(\boldsymbol{r}_{ijk}) &= \sum_s \frac{(\gamma-1)\mu m_p}{k_B} W(r_{s;ijk}/h_s) u_s,
\end{align}
where $h_s,\, m_s,\, u_s,$ and $r_{s;ijk}$ are the $s$-th particle's smoothing
length, mass, specific energy, and distance to the grid point
($r_{s;ijk}=|\boldsymbol{r}_{ijk}-\boldsymbol{r}_s|$, where $\boldsymbol{r}_s$ 
is the position of the $s$-th particle), respectively.
$W(t)$ is an SPH kernel, which is defined as
\begin{equation*}
W(t) \equiv \frac{1}{\pi}
	\begin{cases}
	1-(3/2)t^2+(3/4)t^3 & \mathrm{if}\; 0\le t\le 1 \\
	(2-t)^3/4 & \mathrm{if}\; 1\le t \le 2\\
	0 & \mathrm{otherwise}.
	\end{cases}
\end{equation*}
We assume the specific heat ratio $\gamma = 5/3$ and the mean molecular mass
$\mu = 0.6$. 
The smoothing length of an SPH particle is calculated in the usual way of
the SPH method \citep{HK89,Monaghan92}.

In order to obtain  the mass density at each grid point from dark matter particles 
in each cluster, we use an interpolation technique as in the previous 
SPH method.
The smoothing length of each dark matter particle is set in order that 
each dark matter particle has $32 \pm 3$ neighbors. 

The column density, $\sigma$, is derived from the projection of the total mass 
density, $\rho_{\mathrm{dm}}+\rho_{\mathrm{gas}}$, where $\rho_{\mathrm{dm}}$ 
is dark matter density and $\rho_{\mathrm{gas}}$ is gas density.
The X-ray surface brightness, $\Sigma_X$, is derived from the projection of 
$\rho_{\mathrm{gas}}^2 T^{1/2}$.

\subsection{Statistical Indicators} \label{sec_indicator}
We calculate statistical indicators for each cluster of galaxies to
quantify substructure and perform the KS-tests.
As statistical indicators, we adopt the axial ratio, the $\Mint$, 
the multipole moment power ratio, and the center shift.
Definitions are described below.

\subsubsection{Axial Ratio} \label{sec_ar_def}
The axial ratio is an indicator to show deviation from sphericity of 
a cluster of galaxies \citep{Dutta95, Jing et al.95, Thomas et al.98}.
In a coordinate system whose origin is at the center of mass of the cluster,
the following tensor's eigenvalues are calculated for all dark matter 
particles and SPH particles.
\begin{equation}
I_{ij} = \sum_s m_s x_ix_j.
\end{equation}
These eigenvalues are labeled $\lambda_1,\, \lambda_2,\, \lambda_3$
in decreasing order.
If a cluster is an ellipsoid, its axial ratio is obtained by
$(\lambda_1/\lambda_3)^{1/2}$.

\subsubsection{$\Mint$}
We use $\Mint$ as an indicator of clumpiness of clusters.
We calculate $\Mint$ in a similar way to \citet{Thomas et al.98}.
For each cluster we perform an FoF method for all dark matter particles and 
SPH particles with initial linking length, $l=\bar{n}_c^{-1/3}$, 
where $\bar{n}_c$ is the mean number density of dark matter and SPH particles 
in the cluster.
Next $l$ is gradually lowered until $l$ becomes $(100\bar{n}_c)^{-1/3}$.
This causes the cluster to break up into several subclumps.
At each stage $\Mint$ is defined as
\begin{equation}
\Mint = \frac{m_1+m_2+m_3}{m_1},
\end{equation}
where $m_1\ge m_2 \ge m_3$ are masses of the three largest clumps.
We use the maximum value of $\Mint$ as a measure of 
clumpiness of the cluster.

\subsubsection{Center Shift}
The center shift measures major deviations from symmetry in the cluster mass 
distribution \citep{Jing et al.95, CER96}.
We calculate the center shift for each cluster in a slightly different way
from \citet{Jing et al.95}.
We first search for a peak value, $c_{\mathrm{peak}}$, of $\sigma$ or 
$\Sigma_X$ in the cluster.
The lowest contour level, $c_{\mathrm{lowest}}$, is defined as 
the mean value of $\sigma$ or $\Sigma_X$ at $0.5r_{200}$.
Next the $i$-th contour level is given as $c_i \equiv
c_{\mathrm{peak}}(c_{\mathrm{lowest}}/c_{\mathrm{peak}})^{i/n_{\mathrm{cont}}}$,
where $n_{\mathrm{cont}}$ is the total number of contours.

The center shift, $C$, is defined as
\begin{equation}
 C = \sum_{i=1}^{n_{\mathrm{cont}}} w_i\left\{(x_i-\bar{x})^2+(y_i-\bar{y})^2\right\},
\end{equation}
where $(x_i,y_i)$ is the center of the $i$-th contour, and 
$\bar{x} = \sum_i w_i x_i$, and $\bar{y} = \sum_i w_i y_i$.
The weight of each contour, $w_i$, is proportional to the surface integral
of $\sigma$ or $\Sigma_X$ in the region between this contour and the adjacent 
outer contour.
The center shift shows the emission-weighted dispersion of the centers
of contours.
If a cluster has a a lot of substructure, then the outer contour's center is expected
to shift from $(\bar{x},\bar{y})$ and the center shift becomes large.

\subsubsection{Power Ratio}
The power ratios quantify the shape of projected cluster potentials and
are derived from their multipole expansions \citep[e.g.][]{BT95,BT96,Buote98,Valdarnini99}.

The two-dimensional potential, $\Psi(R,\phi)$, and column density,
$\sigma(R,\phi)$, are related by Poisson's equation,
\begin{equation}
\nabla^2 \Psi(R,\phi) = \sigma(R,\phi), \label{eq_poisson}
\end{equation}
where $R$ and $\phi$ are projected polar coordinates about the center of mass 
of the cluster, and we ignore the constant factor in equation(\ref{eq_poisson}).
The multipole expansion of $\Psi(R,\phi)$ which relies on the 
distribution of the material interior to $R$ \citep[e.g.][]{BT95} is
\begin{align}
\Psi(R,\phi) = -&a_0 \ln \left(\frac{1}{R}\right) - 
	\sum_{m=1}^{\infty} \frac{1}{mR^m} \notag\\
	&\times (a_m\cos m\phi + b_m\sin m\phi), \label{eq_multipole}
\end{align}
where $a_m$ and $b_m$ are defined as follows:
\begin{equation}
a_m = \int_{R'\le R} \sigma(\bx')(R')^m \cos m\phi' \;d^2\bx' \label{eq_am}
\end{equation}
and
\begin{equation}
b_m = \int_{R'\le R} \sigma(\bx')(R')^m \sin m\phi' \;d^2\bx', \label{eq_bm}
\end{equation}
where $\bx' = (R',\phi')$.
The $m$-th moment power, $P_m$, is defined by integration of the square
of the $m$-th term of the multipole expansion, equation (\ref{eq_multipole}), 
over the boundary of a circular aperture of radius $R_{ap}$, 
\begin{equation}
P_m = \frac{1}{2m^2 R_{ap}^{2m}}(a_m^2 + b_m^2)
\end{equation}
for $m>0$, and
\begin{equation}
P_0 = a_0^2
\end{equation}
for $m=0$.

The moment power depends on not only the irregularity of the potential shape 
but also the magnitude of $\sigma$.
Thus we use the moment power ratio, $P_m/P_0$, as an indicator of a cluster's
irregularity.
We define the unit of $R$, the radius in polar coordinates, as $r_{200}$,
so that the power ratio is independent of the size of the cluster.
Since the origin of the polar coordinates is the center of mass of the cluster, 
the dipole moment, $P_1$, is vanished.
The higher order ($m>4$, in this paper) terms are affected by minor
irregularities in cluster shapes, hence we calculate the power ratio for only 
$m=2,3,$ and $4$.
Because the power ratios depend on $R_{ap}$ \citep{BX97,Valdarnini99}, we 
calculate the power ratios for some different values of $R_{ap}$.

We also use the same definition of the power ratio for $\Sigma_X$ 
by replacing $\sigma$ with $\Sigma_X$ in equations \eqref{eq_am} and
\eqref{eq_bm}.

\section{RESULTS} \label{sec_results}
According to the analytical results of \citet{RLT92}, the typical formation 
epoch of galaxy clusters in $\Lambda$CDM is delayed to lower redshift than in OCDM.
This delay clearly appears at low-z ($z\lesssim 0.8$--$0.7$, as shown in 
Fig.\ref{fig_rate}).
We then calculate the indicators described in \S\ref{sec_method}
at $z=0$ and $z=0.5$ and perform the KS-test for each set of indicators 
obtained for two cosmological models.
The KS-test can be used as a statistical test to estimate the ability of
an indicator to be used to discriminate between the model with and without 
$\lambda_0$. 
The result of the KS-test is the probability of the null-hypothesis that
two distributions of the indicator are generated from the same population
\citep{Recipes}.
The significance level of the KS-test adopted in this paper is 5\%.
If the result of the KS-test for an indicator is under this level,
we regard that it is able to distinguish two cosmological models by using 
these indicator.

\subsection{Axial Ratio}
The mean values of the axial ratios are shown in Table \ref{tab_AxialRatio}.
From $z=0.5$ to $z=0$ the axial ratio decreases.
At $z=0$, the axial ratio in $\Lambda$CDM is larger than in OCDM. 
This suggests that the clusters in $\Lambda$CDM have more substructures than 
in OCDM as expected from the analytical prediction. 

\begin{table}
\begin{center}
	\begin{tabular}{lll}
	\tableline\tableline
	redshift & $\Lambda$CDM & OCDM \\\tableline
	$z=0$ & 1.608 & 1.585 \\
	$z=0.5$ & 1.644 & 1.696\\
	\tableline
	\end{tabular}
\end{center}
	\caption{The mean values of the axial ratio.}
	\label{tab_AxialRatio}
\end{table}

The results of the KS-tests for axial ratios are 0.495 for $z=0$ and
0.661 for $z=0.5$.
Both are over the significance level(5\%).
Thus the axial ratio is not useful to distinguish these 
two cosmological models.

\citet{Jing et al.95} performed KS-test for axial ratios, which were 
obtained in a slightly different way to us, and the result of their KS-test 
is under the significance level ($0.13\times 10^{-2}$).
In their calculations of principal axes and axial ratios, they take, 
at first, all particles  within the virial radius of a cluster.
They then calculate new principal axes and axial ratios using particles 
within an ellipsoid with the principal axes and axial ratios just determined.
They repeat the same calculation for the updated ellipsoid until 
the axial ratios converge.
We also calculate axial ratios by their method and perform 
a KS-test on them.
In contrast with their result, the result of our KS-test is over the 
significance level (0.150 for $z=0$ and 0.750 for $z=0.5$) though 
our resolution is similar to theirs.
The reason of this difference is unclear.

\subsection{$\Mint$}
The mean values of $\Mint$ are shown in Table \ref{tab_Mint}.
$\Mint$ in $\Lambda$CDM is larger than that in OCDM at $z=0$.
\begin{table}
\begin{center}
	\begin{tabular}{lll}
	\tableline\tableline
	redshift & $\Lambda$CDM & OCDM \\\tableline
	$z=0$ & 1.289 & 1.188 \\
	$z=0.5$ & 1.284 & 1.189 \\
	\tableline
	\end{tabular}
\end{center}
	\caption{The mean value of the $\Mint$.}
	\label{tab_Mint}
\end{table}

The results of the KS-tests for $\Mint$ are 0.770 for $z=0$ and 0.192
for $z=0.5$.
Both of them are over the significance level.
Again we can not use this indicator to distinguish between 
these two cosmological models. 

The mean values and the results of KS-tests are consistent with 
those of \citet{Thomas et al.98}.

\subsection{Center Shift} \label{sec_shift_result}
The mean values and the standard deviations of the logarithm of the center
shifts are shown in Table \ref{tab_shifts_mean}.
The number of the contours, $n_{\mathrm{cont}}$, is varied from 4 to 8 to 
investigate the effect of the number of the contours.
For large $n_{\mathrm{cont}}$, i.e. a small interval of contour level, 
$C$ is large.
In the case of a small interval, the center shift easily shows small
substructure near to the center of the cluster.

The center shifts for $\sigma$, $C_{\sigma}$, are significantly larger than 
those for $\Sigma_X$, $C_{\Sigma_X}$.
The shapes of the $\Sigma_X$ contours reflect the distribution of the gas.
Gas in a cluster relaxes more quickly than the collisionless dark matter 
\citep{Santabarbara}.
This is the reason why the shapes of the $\Sigma_X$ contours become rounder than 
those of $\sigma$ and the center shifts for $\Sigma_X$ become smaller 
than those for $\sigma$.
\begin{table}
\begin{center}
	\begin{tabular}{lrrr}
\tableline\tableline
Model & $n_{\mathrm{cont}}$ & $\log(C_{\sigma})$ & $\log(C_{\Sigma_X})$\\\tableline
$\Lambda$CDM &&&\\
$(z=0)$& 4 &$ -2.65(\pm 0.72)$ &$ -3.43(\pm 0.78) $\\
& 5 &$ -2.52(\pm 0.69)$ &$ -3.31(\pm 0.82) $\\
& 6 &$ -2.43(\pm 0.67)$ &$ -3.26(\pm 0.80) $\\
& 7 &$ -2.38(\pm 0.64)$ &$ -3.22(\pm 0.80) $\\
& 8 &$ -2.33(\pm 0.63)$ &$ -3.20(\pm 0.77) $\\
OCDM &&&\\
$(z=0)$& 4 &$ -2.80(\pm 0.70)$ &$ -3.64(\pm 0.75) $\\
& 5 &$ -2.72(\pm 0.66)$ &$ -3.56(\pm 0.77) $\\
& 6 &$ -2.62(\pm 0.67)$ &$ -3.49(\pm 0.75) $\\
& 7 &$ -2.59(\pm 0.63)$ &$ -3.47(\pm 0.76) $\\
& 8 &$ -2.54(\pm 0.62)$ &$ -3.46(\pm 0.76) $\\
$\Lambda$CDM &&&\\
$(z=0.5)$& 4 &$ -2.87(\pm 0.58)$ &$ -3.73(\pm 0.67) $\\
& 5 &$ -2.71(\pm 0.60)$ &$ -3.62(\pm 0.64) $\\
& 6 &$ -2.67(\pm 0.57)$ &$ -3.53(\pm 0.64) $\\
& 7 &$ -2.63(\pm 0.55)$ &$ -3.49(\pm 0.67) $\\
& 8 &$ -2.59(\pm 0.55)$ &$ -3.45(\pm 0.64) $\\
OCDM &&&\\
$(z=0.5)$& 4 &$ -2.99(\pm 0.68)$ &$ -3.68(\pm 0.77) $\\
& 5 &$ -2.89(\pm 0.70)$ &$ -3.59(\pm 0.80) $\\
& 6 &$ -2.77(\pm 0.68)$ &$ -3.53(\pm 0.76) $\\
& 7 &$ -2.73(\pm 0.70)$ &$ -3.46(\pm 0.76) $\\
& 8 &$ -2.68(\pm 0.69)$ &$ -3.41(\pm 0.80) $\\
\tableline
\end{tabular}

\end{center}
	\caption{Mean values of logarithm of center shifts}
	\tablecomments{The first column indicates cosmological model and redshift.
		The second column shows the number of contours.
		The third and fourth columns are mean value of $\log(C)$ for
		the column density and the X-ray surface brightness, respectively.}
	\label{tab_shifts_mean}
\end{table}
\begin{figure}
	\plotone{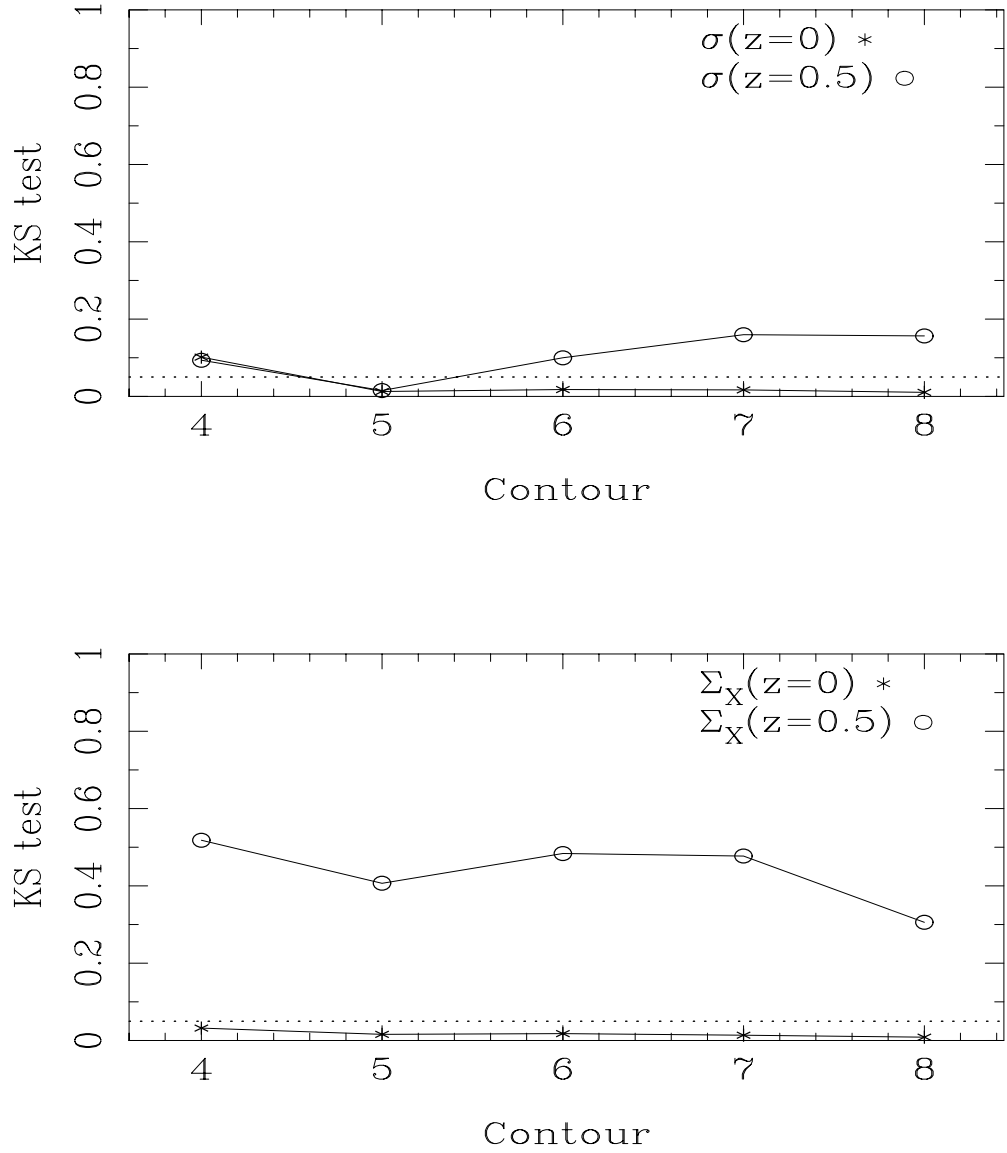}
	\caption{The results of the KS-tests for center shifts as a function of 
		the number of contours.
		Asterisks indicate the results at $z=0$, and circles indicate 
		the results at $z=0.5$.
		The dotted line describes the significance level(5\%)}
	\label{fig_shifts_KS}
\end{figure}

The center shifts in $\Lambda$CDM are larger than those in OCDM
at the same redshift.
This result reflects that the formation epoch of clusters in $\Lambda$CDM
is later than in OCDM. 

Fig.\ref{fig_shifts_KS} shows the result of the KS-tests for the center shifts.
The upper panel describes the results for the center shifts for
the column density and the lower one describes the results for
the X-ray surface brightness.
In Fig.\ref{fig_shifts_KS} the dotted line indicates the significance 
level(5\%), asterisks indicate the results at $z=0$, and circles 
indicate the results at $z=0.5$.
Using the center shifts for $\Sigma_X$ at $z=0$, we can distinguish the two
cosmological models in this range of number of contours.
The effect of contour level interval is not very significant
for the KS-test.

In order to study the effect of the lowest level of contour,
we also calculate the center shifts for $\Sigma_X$ and $\sigma$ in the
two cases that the lowest contour level is the mean value at $0.4r_{200}$ and
$0.7r_{200}$.
For the case of $0.7r_{200}$, the results of the KS-tests are 
over the significance level, except for one case with $C_{\Sigma_X}$ 
(z=0, $n_{\mathrm{cont}}=8$).
For the case of $0.4r_{200}$, the results are similar to
the case of the lowest level at $0.5r_{200}$.
We conclude that the center shift within $\sim 0.5r_{200}$ is 
a useful tool to clarify the presence of a cosmological constant.

\subsection{Power Ratio}
We calculate the power ratios for various $R_{ap}$ (0.4, 0.5, 0.6, 
0.8, and 1.0 times $r_{200}$) in order to study how $R_{ap}$ affects 
the result of the KS-test.
We show the mean values and the standard deviations of the logarithm of
the power ratios in Tables \ref{tab_sigmaPm_mean} and \ref{tab_XPm_mean},
for the column density and the X-ray surface brightness, respectively.
\begin{table}
\begin{center}
	\begin{tabular}{lrrrr}
\tableline\tableline
Model & $R_{ap}$ & $\log(P_2/P_0)$ & $\log(P_3/P_0)$ & $\log(P_4/P_0)$
	\\\tableline
$\Lambda$CDM &&&&\\
$(z=0)$&$ 0.4r_{200}$&$ -2.90(\pm 0.66)$ &$ -4.38(\pm 0.69) $&$ -4.89(\pm 0.75)$\\
&$ 0.5r_{200}$&$ -2.94(\pm 0.63)$ &$ -4.52(\pm 0.71) $&$ -4.91(\pm 0.66)$\\
&$ 0.6r_{200}$&$ -2.98(\pm 0.61)$ &$ -4.61(\pm 0.68) $&$ -4.96(\pm 0.66)$\\
&$ 0.8r_{200}$&$ -3.05(\pm 0.66)$ &$ -4.63(\pm 0.69) $&$ -5.05(\pm 0.72)$\\
&$ 1.0r_{200}$&$ -3.15(\pm 0.68)$ &$ -4.70(\pm 0.67) $&$ -5.20(\pm 0.74)$\\
OCDM &&&&\\
$(z=0)$&$ 0.4r_{200}$&$ -3.03(\pm 0.63)$ &$ -4.58(\pm 0.79) $&$ -5.11(\pm 0.81)$\\
&$ 0.5r_{200}$&$ -3.09(\pm 0.68)$ &$ -4.63(\pm 0.74) $&$ -5.11(\pm 0.80)$\\
&$ 0.6r_{200}$&$ -3.15(\pm 0.76)$ &$ -4.61(\pm 0.76) $&$ -5.10(\pm 0.84)$\\
&$ 0.8r_{200}$&$ -3.20(\pm 0.65)$ &$ -4.72(\pm 0.75) $&$ -5.22(\pm 0.80)$\\
&$ 1.0r_{200}$&$ -3.29(\pm 0.65)$ &$ -4.85(\pm 0.79) $&$ -5.36(\pm 0.76)$\\
$\Lambda$CDM &&&&\\
$(z=0.5)$&$ 0.4r_{200}$&$ -2.93(\pm 0.69)$ &$ -4.33(\pm 0.67) $&$ -4.88(\pm 0.69)$\\
&$ 0.5r_{200}$&$ -2.92(\pm 0.51)$ &$ -4.37(\pm 0.58) $&$ -4.89(\pm 0.65)$\\
&$ 0.6r_{200}$&$ -2.92(\pm 0.50)$ &$ -4.40(\pm 0.57) $&$ -4.85(\pm 0.66)$\\
&$ 0.8r_{200}$&$ -2.91(\pm 0.51)$ &$ -4.35(\pm 0.51) $&$ -4.85(\pm 0.67)$\\
&$ 1.0r_{200}$&$ -2.93(\pm 0.51)$ &$ -4.39(\pm 0.63) $&$ -4.81(\pm 0.64)$\\
OCDM &&&&\\
$(z=0.5)$&$ 0.4r_{200}$&$ -2.76(\pm 0.52)$ &$ -4.29(\pm 0.93) $&$ -4.78(\pm 0.86)$\\
&$ 0.5r_{200}$&$ -2.78(\pm 0.52)$ &$ -4.38(\pm 0.93) $&$ -4.75(\pm 0.80)$\\
&$ 0.6r_{200}$&$ -2.80(\pm 0.54)$ &$ -4.40(\pm 0.90) $&$ -4.79(\pm 0.82)$\\
&$ 0.8r_{200}$&$ -2.87(\pm 0.61)$ &$ -4.43(\pm 0.77) $&$ -4.90(\pm 0.96)$\\
&$ 1.0r_{200}$&$ -3.02(\pm 0.64)$ &$ -4.44(\pm 0.73) $&$ -5.02(\pm 0.94)$\\
\tableline
\end{tabular}

\end{center}
	\caption{Mean values of $\log(P_m/P_0)$ for column density}
	\tablecomments{
		The first column describes cosmological models and redshifts.
		The upper two blocks are the data at $z=0$ and the lower two blocks
		are at $z=0.5$.
		The Second column is aperture radius.
		The third to fifth columns are the mean values of power 
ratios of second, third, and fourth order, respectively.}
	\label{tab_sigmaPm_mean}
\end{table}
\begin{table}
\begin{center}
	\begin{tabular}{lrrrr}
\tableline\tableline
Model & $R_{ap}$ & $\log(P_2/P_0)$ & $\log(P_3/P_0)$ & $\log(P_4/P_0)$
	\\\tableline
$\Lambda$CDM &&&&\\
$(z=0)$&$ 0.4r_{200}$&$ -3.92(\pm 0.96) $&$ -5.86(\pm 1.14) $&$ -6.52(\pm 1.31) $\\
&$ 0.5r_{200}$&$ -4.19(\pm 0.99) $&$ -6.16(\pm 1.14) $&$ -6.81(\pm 1.32) $\\
&$ 0.6r_{200}$&$ -4.40(\pm 1.01) $&$ -6.36(\pm 1.22) $&$ -7.03(\pm 1.34) $\\
&$ 0.8r_{200}$&$ -4.75(\pm 1.05) $&$ -6.66(\pm 1.32) $&$ -7.37(\pm 1.41) $\\
&$ 1.0r_{200}$&$ -5.04(\pm 1.09) $&$ -6.93(\pm 1.40) $&$ -7.67(\pm 1.49) $\\
OCDM &&&&\\
$(z=0)$&$ 0.4r_{200}$&$ -4.40(\pm 0.97) $&$ -6.45(\pm 1.29) $&$ -7.04(\pm 1.42) $\\
&$ 0.5r_{200}$&$ -4.68(\pm 1.03) $&$ -6.70(\pm 1.39) $&$ -7.37(\pm 1.59) $\\
&$ 0.6r_{200}$&$ -4.90(\pm 1.05) $&$ -6.83(\pm 1.40) $&$ -7.51(\pm 1.58) $\\
&$ 0.8r_{200}$&$ -5.23(\pm 1.12) $&$ -7.11(\pm 1.55) $&$ -7.84(\pm 1.70) $\\
&$ 1.0r_{200}$&$ -5.52(\pm 1.15) $&$ -7.38(\pm 1.60) $&$ -8.19(\pm 1.76) $\\
$\Lambda$CDM &&&&\\
$(z=0.5)$&$ 0.4r_{200}$&$ -3.54(\pm 0.83) $&$ -5.41(\pm 1.08) $&$ -5.99(\pm 1.10) $\\
&$ 0.5r_{200}$&$ -3.79(\pm 0.90) $&$ -5.64(\pm 1.07) $&$ -6.30(\pm 1.15) $\\
&$ 0.6r_{200}$&$ -3.97(\pm 0.90) $&$ -5.79(\pm 1.03) $&$ -6.50(\pm 1.21) $\\
&$ 0.8r_{200}$&$ -4.28(\pm 0.97) $&$ -6.04(\pm 1.05) $&$ -6.74(\pm 1.28) $\\
&$ 1.0r_{200}$&$ -4.50(\pm 1.11) $&$ -6.08(\pm 1.31) $&$ -6.76(\pm 1.49) $\\
OCDM &&&&\\
$(z=0.5)$&$ 0.4r_{200}$&$ -3.63(\pm 1.16) $&$ -5.35(\pm 1.53) $&$ -6.03(\pm 1.64) $\\
&$ 0.5r_{200}$&$ -3.87(\pm 1.20) $&$ -5.59(\pm 1.59) $&$ -6.26(\pm 1.75) $\\
&$ 0.6r_{200}$&$ -4.04(\pm 1.21) $&$ -5.83(\pm 1.66) $&$ -6.45(\pm 1.79) $\\
&$ 0.8r_{200}$&$ -4.29(\pm 1.35) $&$ -6.05(\pm 1.77) $&$ -6.71(\pm 2.05) $\\
&$ 1.0r_{200}$&$ -4.59(\pm 1.40) $&$ -6.30(\pm 1.87) $&$ -6.98(\pm 2.08) $\\
\tableline
\end{tabular}

\end{center}
	\caption{Mean values of $\log(P_m/P_0)$ for X-ray surface brightness}
	\label{tab_XPm_mean}
\end{table}
For small $R_{ap}$, the power ratios are large as shown in Table 
\ref{tab_sigmaPm_mean} and \ref{tab_XPm_mean}.
The most likely explanation of this property is that the power ratios for
small $R_{ap}$ readily reflect the small substructures around the cluster 
center. 

\begin{table}
\begin{center}
	\begin{tabular}{lrrrr}
\tableline\tableline
Model & $R_{ap}$ & $\log(P_2/P_0)$ & $\log(P_3/P_0)$ & $\log(P_4/P_0)$
	\\\tableline
$\Lambda$CDM &&&&\\
$(z=0)$&$ 0.4h^{-1}\mathrm{Mpc}$&$ -5.65(\pm 0.85) $&$ -7.66(\pm 1.14) $&$ -8.22(\pm 1.23) $\\
&$ 0.6h^{-1}\mathrm{Mpc}$&$ -6.03(\pm 0.93) $&$ -7.97(\pm 1.23) $&$ -8.62(\pm 1.32) $\\
&$ 0.8h^{-1}\mathrm{Mpc}$&$ -6.39(\pm 0.98) $&$ -8.39(\pm 1.14) $&$ -8.98(\pm 1.31) $\\
&$ 1.0h^{-1}\mathrm{Mpc}$&$ -6.68(\pm 1.02) $&$ -8.63(\pm 1.24) $&$ -9.26(\pm 1.34) $\\
&$ 1.2h^{-1}\mathrm{Mpc}$&$ -6.90(\pm 1.05) $&$ -8.81(\pm 1.31) $&$ -9.52(\pm 1.45) $\\
OCDM &&&&\\
$(z=0)$&$ 0.4h^{-1}\mathrm{Mpc}$&$ -5.96(\pm 0.89) $&$ -8.05(\pm 1.20) $&$ -8.57(\pm 1.29) $\\
&$ 0.6h^{-1}\mathrm{Mpc}$&$ -6.47(\pm 0.96) $&$ -8.55(\pm 1.21) $&$ -9.09(\pm 1.40) $\\
&$ 0.8h^{-1}\mathrm{Mpc}$&$ -6.83(\pm 1.05) $&$ -8.85(\pm 1.39) $&$ -9.48(\pm 1.58) $\\
&$ 1.0h^{-1}\mathrm{Mpc}$&$ -7.12(\pm 1.08) $&$ -9.07(\pm 1.47) $&$ -9.74(\pm 1.64) $\\
&$ 1.2h^{-1}\mathrm{Mpc}$&$ -7.35(\pm 1.12) $&$ -9.25(\pm 1.56) $&$ -9.95(\pm 1.70) $\\
$\Lambda$CDM &&&&\\
$(z=0.5)$&$ 0.4h^{-1}\mathrm{Mpc}$&$ -5.35(\pm 1.04) $&$ -7.11(\pm 1.46) $&$ -7.70(\pm 1.43) $\\
&$ 0.6h^{-1}\mathrm{Mpc}$&$ -5.78(\pm 1.20) $&$ -7.57(\pm 1.54) $&$ -8.22(\pm 1.69) $\\
&$ 0.8h^{-1}\mathrm{Mpc}$&$ -6.13(\pm 1.28) $&$ -7.99(\pm 1.40) $&$ -8.58(\pm 1.80) $\\
&$ 1.0h^{-1}\mathrm{Mpc}$&$ -6.36(\pm 1.33) $&$ -8.17(\pm 1.48) $&$ -8.68(\pm 1.81) $\\
&$ 1.2h^{-1}\mathrm{Mpc}$&$ -6.58(\pm 1.43) $&$ -8.15(\pm 1.62) $&$ -8.80(\pm 1.91) $\\
OCDM &&&&\\
$(z=0.5)$&$ 0.4h^{-1}\mathrm{Mpc}$&$ -5.58(\pm 1.21) $&$ -7.28(\pm 1.64) $&$ -7.95(\pm 1.74) $\\
&$ 0.6h^{-1}\mathrm{Mpc}$&$ -6.05(\pm 1.26) $&$ -7.89(\pm 1.59) $&$ -8.48(\pm 1.85) $\\
&$ 0.8h^{-1}\mathrm{Mpc}$&$ -6.39(\pm 1.35) $&$ -8.12(\pm 1.75) $&$ -8.74(\pm 2.03) $\\
&$ 1.0h^{-1}\mathrm{Mpc}$&$ -6.70(\pm 1.38) $&$ -8.44(\pm 1.78) $&$ -9.14(\pm 2.12) $\\
&$ 1.2h^{-1}\mathrm{Mpc}$&$ -6.94(\pm 1.39) $&$ -8.67(\pm 1.80) $&$ -9.34(\pm 1.99) $\\
\tableline
\end{tabular}

\end{center}
	\caption{Mean values of $\log(P_m/P_0)$ with constant $R_{ap}$ 
	for X-ray surface brightness.} 
	\label{tab_XPmMpc_mean}
\end{table}
We also calculate the power ratios for constant $R_{ap}$, e.g.
$0.4h^{-1}\mathrm{Mpc}$, and $0.6h^{-1}\mathrm{Mpc}$.
As shown in Table \ref{tab_XPmMpc_mean}, the power ratios for the constant 
$R_{ap}$ are smaller than the proportional $R_{ap}$ case because
the constant $R_{ap}$ is less sensitive to the small substructures
in small clusters than is the proportional $R_{ap}$.
In the constant $R_{ap}$ case, the mean values and the standard deviations 
of the power ratios agree with \citet{Valdarnini99}.

From $z=0.5$ to $z=0$ the power ratios become smaller in both models.
This is due to the fact that the fraction of relaxed clusters increases with time.
Like the other indicators, the power ratios in $\Lambda$CDM are
larger than those in OCDM at the same redshift as expected. 

\begin{figure}
	\plotone{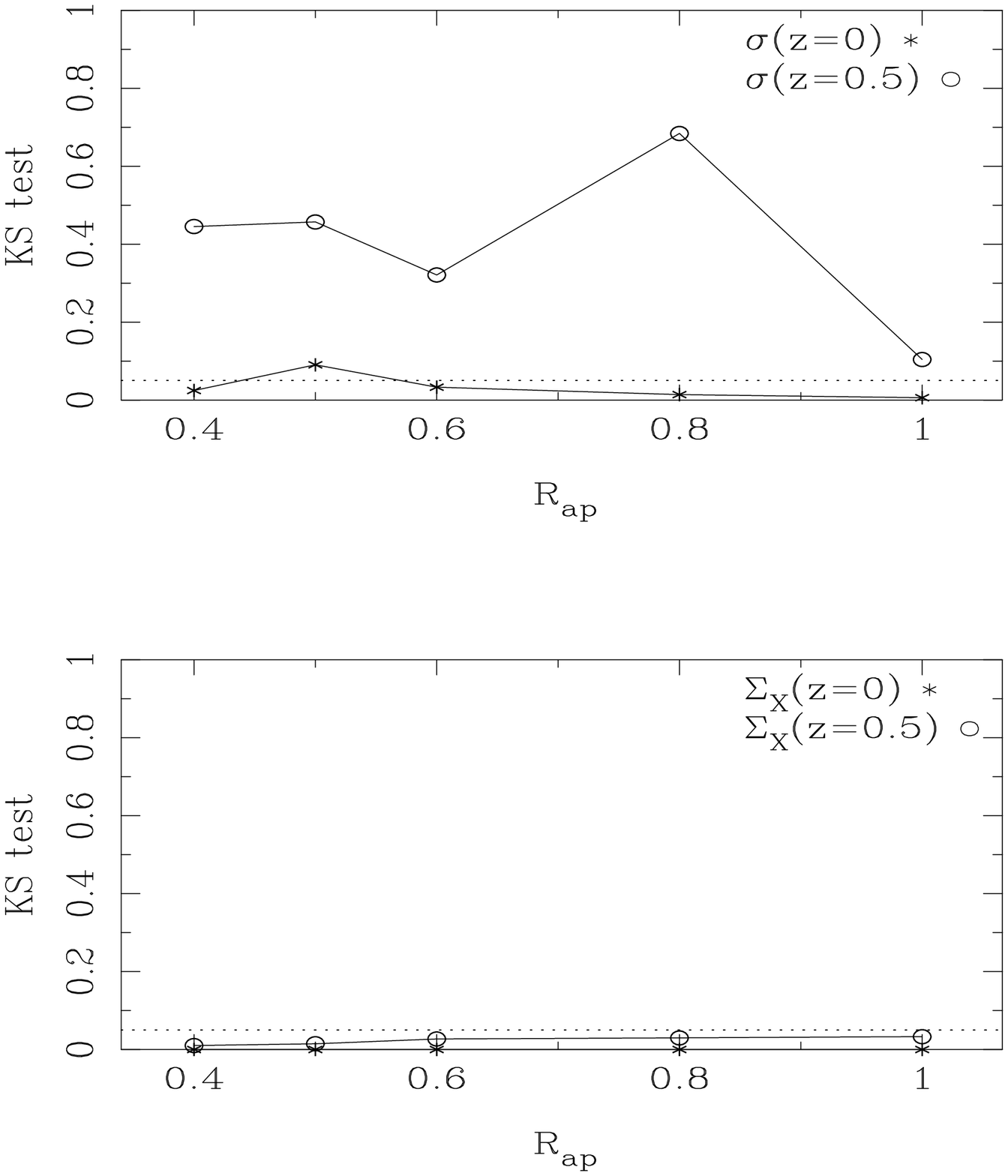}
	\caption{The same as Fig.\ref{fig_shifts_KS} but for $P_2/P_0$.}
	\label{fig_P2_KS}
\end{figure}

\begin{figure}
	\plotone{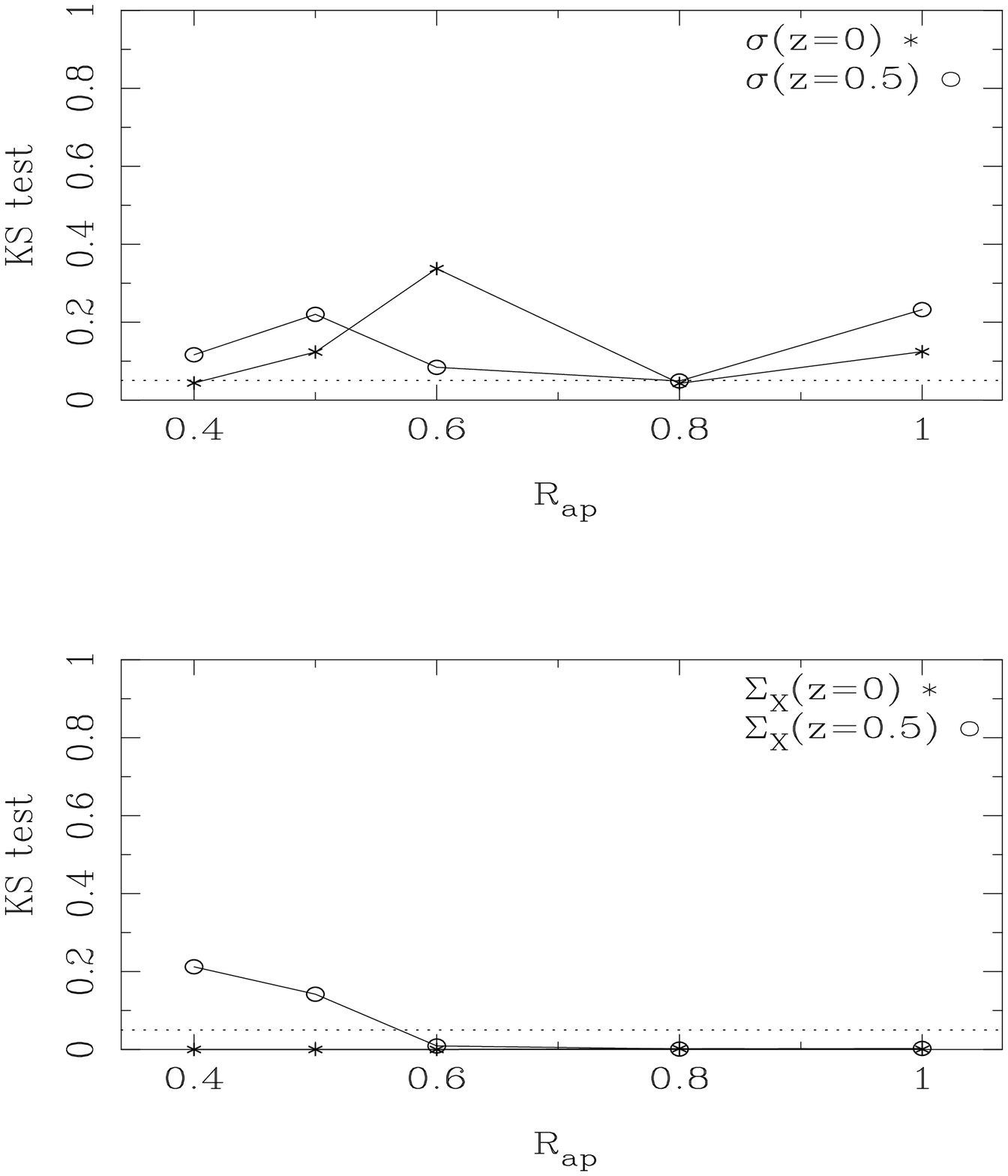}
	\caption{The same as Fig.\ref{fig_shifts_KS} but for $P_3/P_0$.}
	\label{fig_P3_KS}
\end{figure}

\begin{figure}
	\plotone{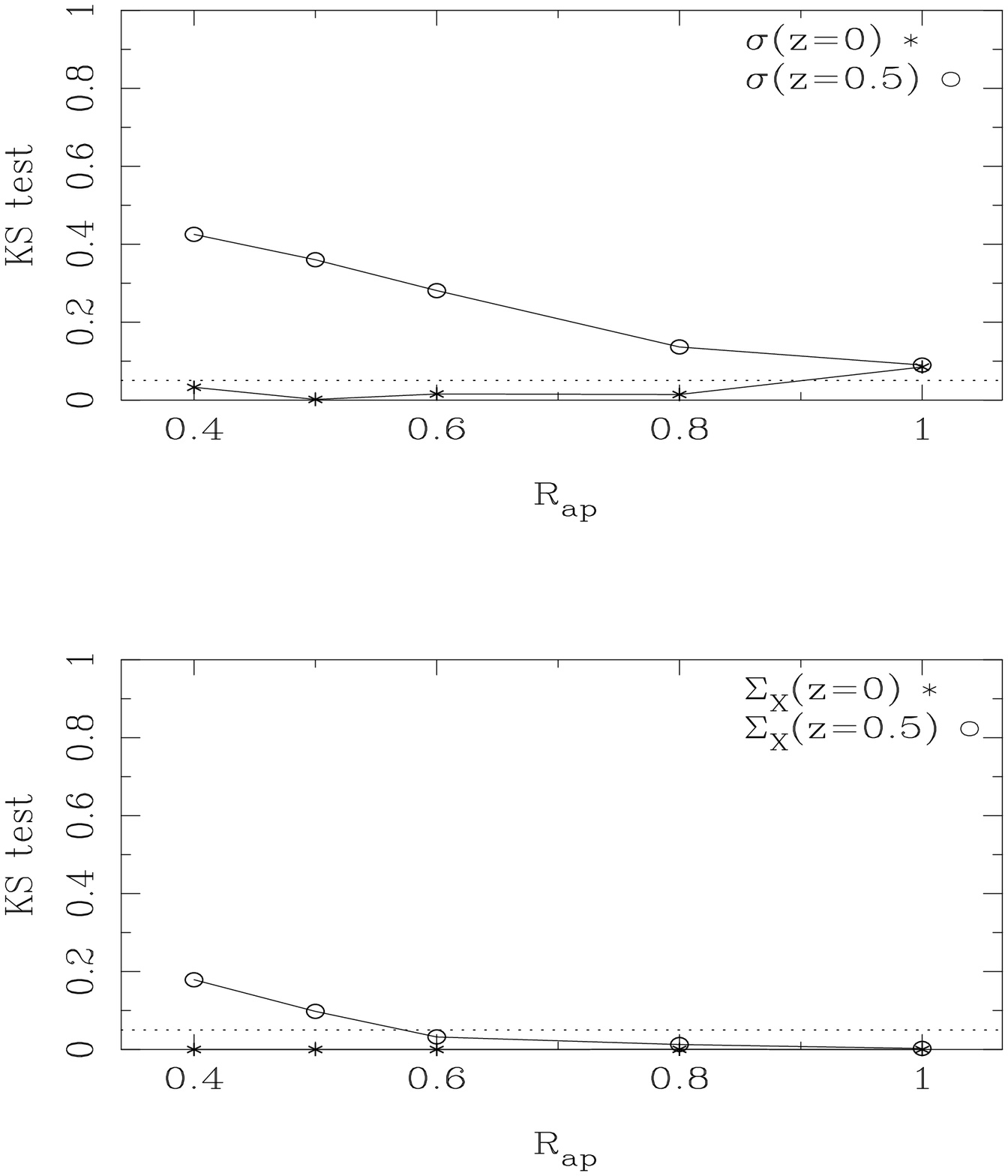}
	\caption{The same as Fig.\ref{fig_shifts_KS} but for $P_4/P_0$.}
	\label{fig_P4_KS}
\end{figure}

Table \ref{tab_SPmks} shows the results of the KS-tests for the power ratios 
for the column density, and Table \ref{tab_XPmks} shows those for the X-ray
surface brightness.
The figures with asterisks in these tables mean that they are under the 
significance level (5\%).
Figures \ref{fig_P2_KS}--\ref{fig_P4_KS} show the results of the KS-tests
summarized in Tables \ref{tab_SPmks} and \ref{tab_XPmks}.
At $z=0$, we can distinguish the two cosmological models by the power ratios for
$\Sigma_X$ for all $R_{ap}$.
Using the power ratios for $\sigma$ for most $R_{ap}$ except for $P_3/P_0$,
we can also distinguish the two cosmological models.
\begin{table}
\begin{center}
	\begin{tabular}{lrrr}
\tableline\tableline
$P_m/P_0$ & $R_{ap}$ & $z=0$ & $z=0.5$\\\tableline

$P_{2}/P_0$&&&\\
& $ 0.4r_{200}$ & {2.43e-02}* &{4.45e-01} \\
& $ 0.5r_{200}$ & {9.03e-02} &{4.57e-01} \\
& $ 0.6r_{200}$ & {3.30e-02}* &{3.21e-01} \\
& $ 0.8r_{200}$ & {1.38e-02}* &{6.84e-01} \\
& $ 1.0r_{200}$ & {5.79e-03}* &{1.04e-01} \\
$P_{3}/P_0$&&&\\
& $ 0.4r_{200}$ & {4.38e-02}* &{1.16e-01} \\
& $ 0.5r_{200}$ & {1.23e-01} &{2.20e-01} \\
& $ 0.6r_{200}$ & {3.37e-01} &{8.40e-02} \\
& $ 0.8r_{200}$ & {4.28e-02}* &{4.87e-02}* \\
& $ 1.0r_{200}$ & {1.24e-01} &{2.32e-01} \\
$P_{4}/P_0$&&&\\
& $ 0.4r_{200}$ & {3.25e-02}* &{4.25e-01} \\
& $ 0.5r_{200}$ & {1.40e-03}* &{3.60e-01} \\
& $ 0.6r_{200}$ & {1.53e-02}* &{2.81e-01} \\
& $ 0.8r_{200}$ & {1.43e-02}* &{1.36e-01} \\
& $ 1.0r_{200}$ & {8.48e-02} &{8.97e-02} \\
\tableline
\end{tabular}

\end{center}
	\caption{The results of the KS-tests for power ratios for the column density.}
	\label{tab_SPmks}
\end{table}
\begin{table}
\begin{center}
	\begin{tabular}{lrrr}
\tableline\tableline
$P_m/P_0$ & $R_{ap}$ & $z=0$ & $z=0.5$\\\tableline

$P_{2}/P_0$&&&\\
& $ 0.4r_{200}$ & {4.19e-08}* &{1.03e-02}* \\
& $ 0.5r_{200}$ & {7.20e-08}* &{1.47e-02}* \\
& $ 0.6r_{200}$ & {1.33e-06}* &{2.70e-02}* \\
& $ 0.8r_{200}$ & {1.47e-06}* &{2.99e-02}* \\
& $ 1.0r_{200}$ & {2.56e-06}* &{3.30e-02}* \\
$P_{3}/P_0$&&&\\
& $ 0.4r_{200}$ & {6.75e-06}* &{2.12e-01} \\
& $ 0.5r_{200}$ & {3.51e-05}* &{1.42e-01} \\
& $ 0.6r_{200}$ & {4.43e-05}* &{9.17e-03}* \\
& $ 0.8r_{200}$ & {2.24e-03}* &{1.15e-03}* \\
& $ 1.0r_{200}$ & {7.99e-04}* &{2.43e-03}* \\
$P_{4}/P_0$&&&\\
& $ 0.4r_{200}$ & {2.06e-06}* &{1.79e-01} \\
& $ 0.5r_{200}$ & {6.83e-05}* &{9.77e-02} \\
& $ 0.6r_{200}$ & {4.72e-05}* &{3.22e-02}* \\
& $ 0.8r_{200}$ & {2.52e-04}* &{1.28e-02}* \\
& $ 1.0r_{200}$ & {7.56e-05}* &{2.68e-03}* \\
\tableline
\end{tabular}

\end{center}
	\caption{The results of the KS-tests for power ratios for the X-ray surface brightness.}
	\label{tab_XPmks}
\end{table}

\section{DISCUSSION AND CONCLUSIONS} \label{sec_discussion}
We have investigated morphologies of the galaxy clusters in 
$\Lambda$CDM and OCDM at $z = 0$ and $z = 0.5$ using large 
hydrodynamic simulations.  
For clusters in each model we have calculated the axial ratios, $\Mint$, 
center shifts, and multipole moment power ratios as statistical 
indicators that quantify the irregularity of cluster morphologies.

The power ratios and the center shifts are calculated for  
the projected density ($\sigma$) as well as the X-ray surface brightness
($\Sigma_X$).  
For $\sigma$, both indicators show a larger value than those for $\Sigma_X$, 
because the relaxation time scale of the collisionless particles is much 
longer than that of the collisional gas particles  
\citep{Santabarbara,Valdarnini99}.

At $z=0$ all mean values of statistical indicators in $\Lambda$CDM show
larger values than those in OCDM.
These large values, which indicate large irregularity of the clusters, 
suggest more recent formation of the clusters in $\Lambda$CDM as
expected from the analytic prediction \citep{RLT92} 
and the previous numerical studies \citep{M95,CER96,BX97}.

We use KS-tests to estimate the ability of the indicators to distinguish
between two cosmological models.
From the results of these KS-tests, the distributions of axial ratios in
the two 
cosmological models are indistinguishable.
The distributions of $\Mint$ in $\Lambda$CDM and OCDM are also similar.
Using the center shifts and the power ratios for $\Sigma_X$ at $z=0$, we can
distinguish the two cosmological models.
It is possible to discriminate between $\Lambda$CDM and OCDM using the 
center shifts, $P_2/P_0$, and $P_4/P_0$ for $\sigma$ at $z=0$.
At $z=0.5$ we can distinguish the two cosmological models by the power ratios 
for $\Sigma_X$, but not by the center shifts for $\Sigma_X$ or 
both the power ratios and the center shifts for $\sigma$.

Using the power ratios for $\Sigma_X$ we can distinguish between $\Lambda$CDM
and OCDM better than using those for $\sigma$ as shown in Table
\ref{tab_SPmks} and \ref{tab_XPmks}.
This may be due to the fact that the relaxation time scale of dark matter 
that dominates $\sigma$ is longer than that of the gas on which 
$\Sigma_X$ mainly depends, as described above.
Since $\sigma$ does not settle rapidly after the cluster formation epoch, 
the power ratios for $\sigma$ are relatively insensitive to
the difference of the cluster formation epochs.

We also note that the power ratios for $\Sigma_X$ depend more on 
irregularities in high density regions (i.e. in the central region) 
than those for $\sigma$,
because $\Sigma_X$ is proportional to square of gas density (and square root of
temperature) as $\rho_{\mathrm{gas}}^2T^{1/2}$.
Since small substructures which appear in the central region are
expected to be erased more easily than large scale substructures,
the power ratios for $\Sigma_X$ are probably more sensitive to the cluster 
formation epoch than those for $\sigma$.
In order to clarify this point we calculate power ratios for $\sigma^2$, 
and then perform KS-tests.
The results for $\sigma^2$ are better than those for $\sigma$, but worse 
than those for $\Sigma_X$ reflecting the difference in relaxation
time-scale. 
From these facts, we suggest that indicators that focus on the central 
regions of clusters are suitable to distinguish between $\Lambda$CDM
and OCDM. 

Some previous studies \citep{Jing et al.95,CER96,BX97} compared 
$\Lambda$CDM and OCDM by using pure N-body simulations in which they 
assume that $\rho_{\mathrm{gas}}$ is proportional to $\rho_{\mathrm{dm}}$.
For clusters which are not well relaxed, however, this assumption may not 
be valid.
Using the power ratios for $\Sigma_X$ \citet{BX97} could not 
distinguish $\Lambda$CDM and OCDM, while we can. 
On the other hand, \citet{Jing et al.95} and \citet{CER96} showed that 
$\Lambda$CDM and OCDM are distinguishable by using center shifts, and 
we also reach the same conclusion.
These results suggest that the power ratios are more sensitive to the 
existence of gas than the center shifts.

We shall discuss differences in the properties of these two indicators.
The center shifts show a large value when there are substructures
whose sizes are large enough to shift a center of contour from 
the inner contour to the outer one.
Since the center shifts are sensitive to large size substructures that can 
be detected in the distributions of both gas and dark matter in the cluster, 
the center shifts for the simulations with and without gas are expected 
to be similar.
In contrast, the power ratios are more sensitive to small size substructures 
than the center shifts and such small size substructures are sometimes 
detected in only the dark matter distribution since corresponding gas 
structures are already relaxed.
Hence the difference between the power ratios for the simulations with and 
without gas is expected to be larger than that of the center shifts.
In order to confirm this explanation, we calculate correlation coefficients 
of the power ratios for $\Sigma_X$ and $\sigma$, and those of center shifts 
for $\Sigma_X$ and $\sigma$.
As expected, the correlation coefficients of the power ratios 
tend to be smaller than those of the center shifts.
Therefore it is important to include a gas component in simulations 
in order to distinguish the two cosmological models using the power ratios.

Although  some authors \citep{Evrard93,M95} employed gas particles, 
the number of clusters was too small (8 clusters for each model).
We use much larger size simulations ($L_{\mathrm{box}}=150h^{-1}
\mathrm{Mpc}$) than the previous studies in order to obtain a sufficient
number of clusters to perform statistical tests (70--80 clusters at z=0).
The KS-test is significant when the number of samples is larger than 20
\citep{Recipes}.

\citet{Valdarnini99} carried out hydrodynamic simulations for a sufficient 
number of clusters to perform statistical tests.
By comparing their simulated clusters with ROSAT clusters, they showed that 
power ratios for gas in SCDM universes are in much better agreement 
with the ROSAT data than those for $\Lambda$CDM.
Although this result is quite interesting, in order to give more 
conclusive results we should wait Chandra and XMM data which will be 
high resolution and high S/N and much higher resolution simulations with 
additional important physical processes such as radiative cooling, 
SNe feedback, and so on.  

By comparison with the power ratios in $\Lambda$CDM obtained by 
\citet{Valdarnini99}, those obtained in this paper are systematically 
large.
This difference results from the fact that the $R_{ap}$ for 
our power ratios is proportional to $r_{200}$, while $R_{ap}$ 
in their paper was constant.
Thus, our power ratios are more sensitive to small-scale irregularity for 
small clusters than theirs.
Mean values of power ratios and its standard deviations in our simulations 
agree well with their results for the constant $R_{ap}$ case.

\citet{Schuecker01} showed that the fraction of clusters of galaxies with 
substructures is almost 50\% for REFLEX+BCS clusters after correction for 
several systematic effects.
They also suggested that the fraction of clusters with substructures 
depends on the number density of the clusters.
Then this large fraction may be explained by the fact that a large 
fraction ($\sim 54\%$ for $z\leq 0.1$) of rich clusters appear to belong 
to superclusters \citep{supercluster}.
On the other hand,  the size of simulation box of our and 
previous simulations may not be large enough for the realization 
of large superclusters, such as the Shapley supercluster, since the Shapley 
supercluster is expected to be formed from a 3.5$\sigma$ perturbation with 
a size of $\sim$30Mpc (Ettori, Fabian, \& White 1997).
This may be the reason why the previous simulations of open and flat
$\Lambda$ universes do not agree with the ROSAT X-ray observation data. 
We will simulate initial conditions corresponding to the local universe to 
examine this possibility.

Our conclusion is as follows.
The power ratios and the center shifts for $\Sigma_X$ can be used to 
distinguish between $\Lambda$CDM and OCDM.
Since the formation histories of clusters in $\Lambda$CDM are similar 
to those in OCDM compared with the high-density universe, 
these indicators will be powerful probes to test the cosmological models 
in the non-linear regime.

\acknowledgments
We are grateful to Vincent Eke for helpful advice. We also thank the
anonymous referee for valuable comments. 
Numerical computation in this work was carried out on the SGI Origin 2000 
at the Division of Physics, Graduate School of Science, Hokkaido University, 
and on the VPP300/16R and VPP5000 at the Astronomical Data Analysis Center of 
National Astronomical Observatory, Japan.

\end{document}